\documentclass[11pt,twoside]{article}
\usepackage{asp2014}

\aspSuppressVolSlug
\resetcounters

\begin{document}

\title{New and improved software for data processing at HartRAO}

\author{Pfesesani~V.~van~Zyl}
\affil{South African Radio Astronomy Observatory, Krugersdorp, Gauteng, South Africa; \email{pvanzyl@sarao.ac.za}}

\paperauthor{Pfesesani~V.~van~Zyl}{pvanzyl@sarao.ac.za}{0000-0002-7510-6366}{SARAO}{Fundamental astronomy}{Krugersdorp}{Gauteng}{1740}{South Africa}





\begin{abstract}
The Hartebeesthoek Radio Astronomy Observatory (HartRAO) has been processing its data using {\tt LINES}, a {\tt Fortran}-based program developed in 1989. However, due to the lack of adequate software updates over recent years, the program has become difficult to work with, sighting problems ranging from compatibility issues with newer operating systems to maintenance issues from using a generally unfamiliar programming language. This work presents {\tt DRAN}, a new software package for the reduction and analysis of HartRAO single-dish continuum data. {\tt DRAN}'s main functionality is based on {\tt LINES}, however, it was developed using {\tt Python} and offers a variety of advanced features that include automated data flagging, outlier detection, flux calibration, and time-series analysis which were not previously available in {\tt LINES}. The objective of this project was to produce a standard user-friendly software package for the observatory that produces timeously calibrated data, drift scan images, and supporting documentation for users of HartRAO continuum data. 
\end{abstract}

\section{Introduction}

For the past 30 years, the Hartebeesthoek Radio Astronomy Observatory (HartRAO)\footnote{http://www.hartrao.ac.za} has been processing single-dish radio observations from the 26-m HartRAO telescope\footnote{http://www.hartrao.ac.za/hh26m$\_$factsfile.html} using the spectroscopy analysis program ({\tt LINES}), a software program developed by Dr. M.J. Gaylard as part of his Ph.D. project \citep{gaylard_radio_1989}. The program was developed in {\tt Fortran 77}, a powerful general-purpose programming language developed to optimize numerically intensive computations that require high-performance and fast computational speeds. {\tt LINES} was originally created as part of a suite of programs that had to initiate and carry out single-dish spectral line observations on an HP1000 mini-computer used to operate the 26-m telescope at HartRAO. However, it later got adapted to also process drift-scan observations of radio continuum sources. Since 2007, the  {\tt LINES} program has not been regularly maintained, which has raised several critical issues:

	- The program has become difficult to install on modern systems (installation requires expert assistance).
	
	-  External scripts are required to run automated data reduction, batch processing.
	 
   - The program does not work with {\tt FITS}\footnote{ https://fits.gsfc.nasa.gov/fits$\_$libraries.html} files (observing files need to be converted to {\tt ASCII} before processing; also done by an external script).
   
	- Extracting results from the program is often time-consuming; a user is required to manually copy results from the output screen.
	
    - There are tools to perform basic data reduction, but no additional tools are in place to perform further data analysis. This hinders the program's ability to provide the level of reporting required by the observatory's changing science needs.
 
	- The program requires a substantial investment to upgrade as {\tt Fortran 77} is very old.

\noindent
\\
Research software needs to be sustainable in order to understand, replicate, reproduce, and build upon existing research or conduct new research effectively. Working with old, unsupported programs  adds  further complications as it can be challenging and time-consuming to learn an old language. 

To navigate these problems, the drift-scan data reduction and analysis software ({\tt DRAN}) was created. {\tt DRAN} is the new single-dish data reduction software for HartRAO continuum data written in {\tt Python}, a modern programing language widely used in the astronomical community. This program was developed with the aim to create a centralized software package for students and researchers to reduce, calibrate, and analyze HartRAO data. The program should also be able to 
run automated pipelines (set of specialized operations within the larger program workflow) for processing of single-dish calibration and monitoring observations at a daily cadence. {\tt DRAN} is designed with a user-friendly command-line and graphical user interface-based program that processes single-dish continuum data in single-file and batch mode. The program is easy to use and makes data processing easier by automating many of the user tasks. {\tt DRAN} also offers a range of new features and improvements, including automated data flagging, fitting, and outlier detection, and time-series analysis.


\section{Software technology stack}
Many of the basic operations performed by the {\tt DRAN}  program leverage functionality from existing well-known {\tt Python} scientific libraries. {\tt  Numpy}, {\tt Scipy}, {\tt Pyfits}, and {\tt Astropy} provide  mathematical and data conversions. {\tt Matplotlib} and {\tt Pyqt5} provide the platforms for the visualizations and development of the new graphical user interface. Data manipulation of the reduced data utilizes  {\tt Pandas}, and output from the data reduction process is stored in an {\tt SQLite} database. All program documentation is provided by {\tt SPHINX}.

\section{Data processing pipeline}
{\tt DRAN}'s program design is centered around an object-oriented model (OOM). The OOM uses object instances to call methods to execute specific chains of events. Figure~\ref{ex_fig1} shows the basic outline of the {\tt DRAN} workflow outline. The pipeline follows three basic paths.
Firstly, the data is cleansed. This stage of the data reduction process eliminates any misformed or incomplete observations. Any files that do not follow the expected {\tt FITS} format for each specific frequency get removed at this level of the process. Secondly, the data is prepared. Here, the calibration steps are initiated, converting source counts to temperature units, as well as calibrating the atmospheric effects that could affect the observed signal. Lastly, the data is then processed. The data is checked for outliers which when detected are removed from the data using an automated {\tt clean\_RFI} script that isolates all data points above and below a specified threshold (currently set at 2.7 times the rms-error of the splined fit applied to the entire drift-scan). Once the 
outliers are removed, a low-order polynomial is fit to the baseline to remove any drift in the baseline that may have occurred during observation. The peak of the drift-scan is then fitted using a second-order polynomial and the results are subsequently stored in the SQLite database. Plots generated from the intermediate fitting steps as well as the final plot of the peak fit are then also stored as PNG files in the plots folder to complete the pipeline cycle. This process can be run manually or automatically for both single-file and batch mode operations. An added feature that simplifies this process is the automated fitting routine. This routine runs the {\tt DRAN} pipeline in automatic mode, with the program making all the decisions on whether or not to process a file, and where along the drift-scan main beam to perform the baseline and peak fits. This is a feature that {\tt LINES} did not have and has proven to be extremely beneficial to the overall pipeline strategy, especially when running batch analysis.

\articlefigure{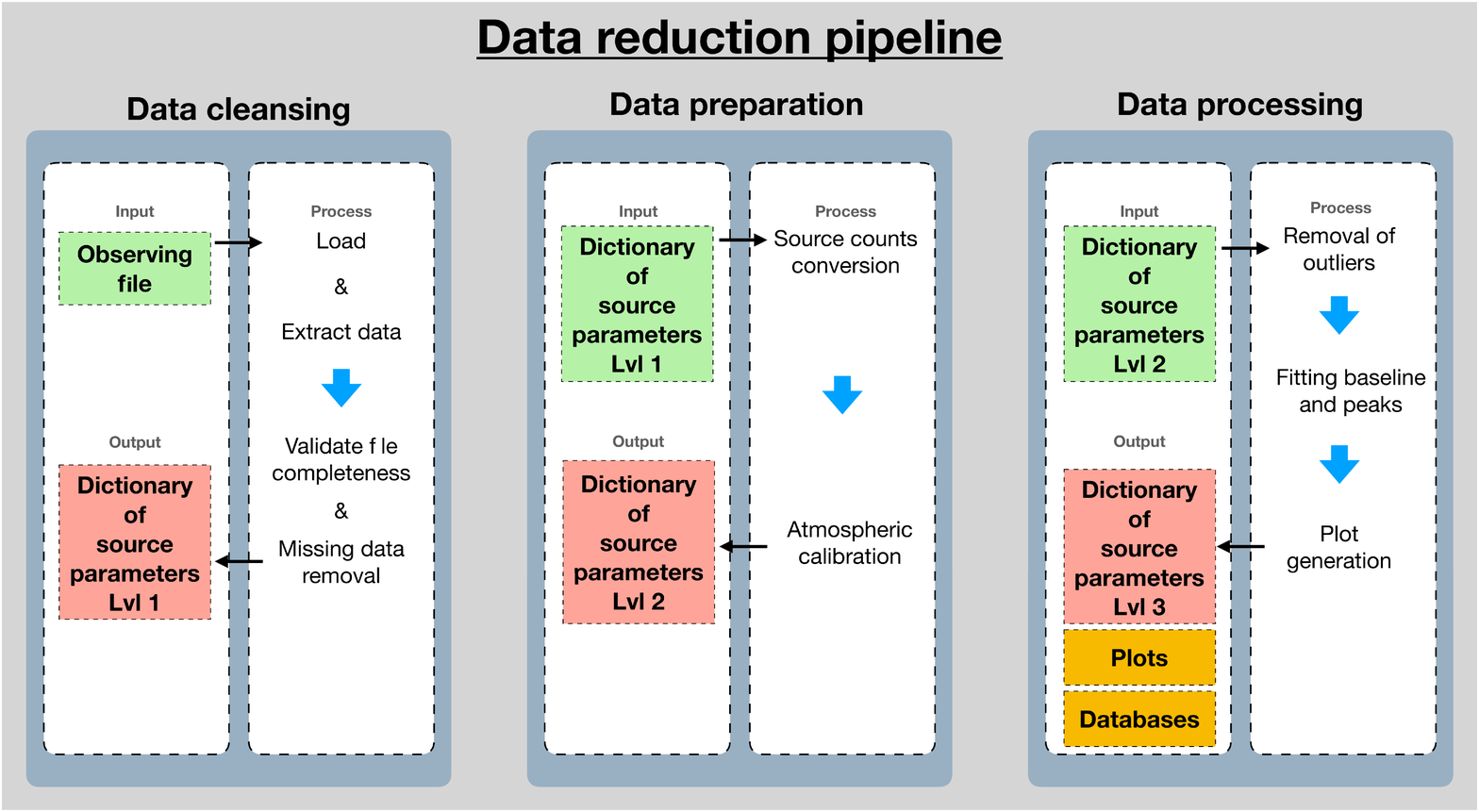}{ex_fig1}{{\tt DRAN} workflow outline.}

\subsection{The new interface}

{\tt DRAN}  consists of two new basic interfaces, the command-line interface, and the graphical user interface. Figure~\ref{ex_fig2} shows a snapshot of the new graphical user interface (GUI). The GUI can be loaded from the {\tt DRAN} command-line interface and is a multi-functional tool that provides an easy-to-navigate, interactive space where data can be viewed and modified. The GUI can follow the same pipeline discussed above, and also comes with an additional set of interfaces for data reduction of drift-scans, time-series analysis, and plot visualization. The layout of the GUI is set up into 3 main grids, the INFO GRID, which displays basic information on the observed object. In the INFO GRID, there's also a tab button to toggle between the properties and the fit window, the latter lets the user perform fitting routines on the data by interacting with the CANVAS GRID. The DISPLAY GRID lets the user toggle between different views of the target objects' drift-scans. And lastly, the CANVAS GRID shows the current drift-scan under investigation, all residuals from fitting the data are shown in the Residuals window at the foot of the main canvas. 

\articlefigure{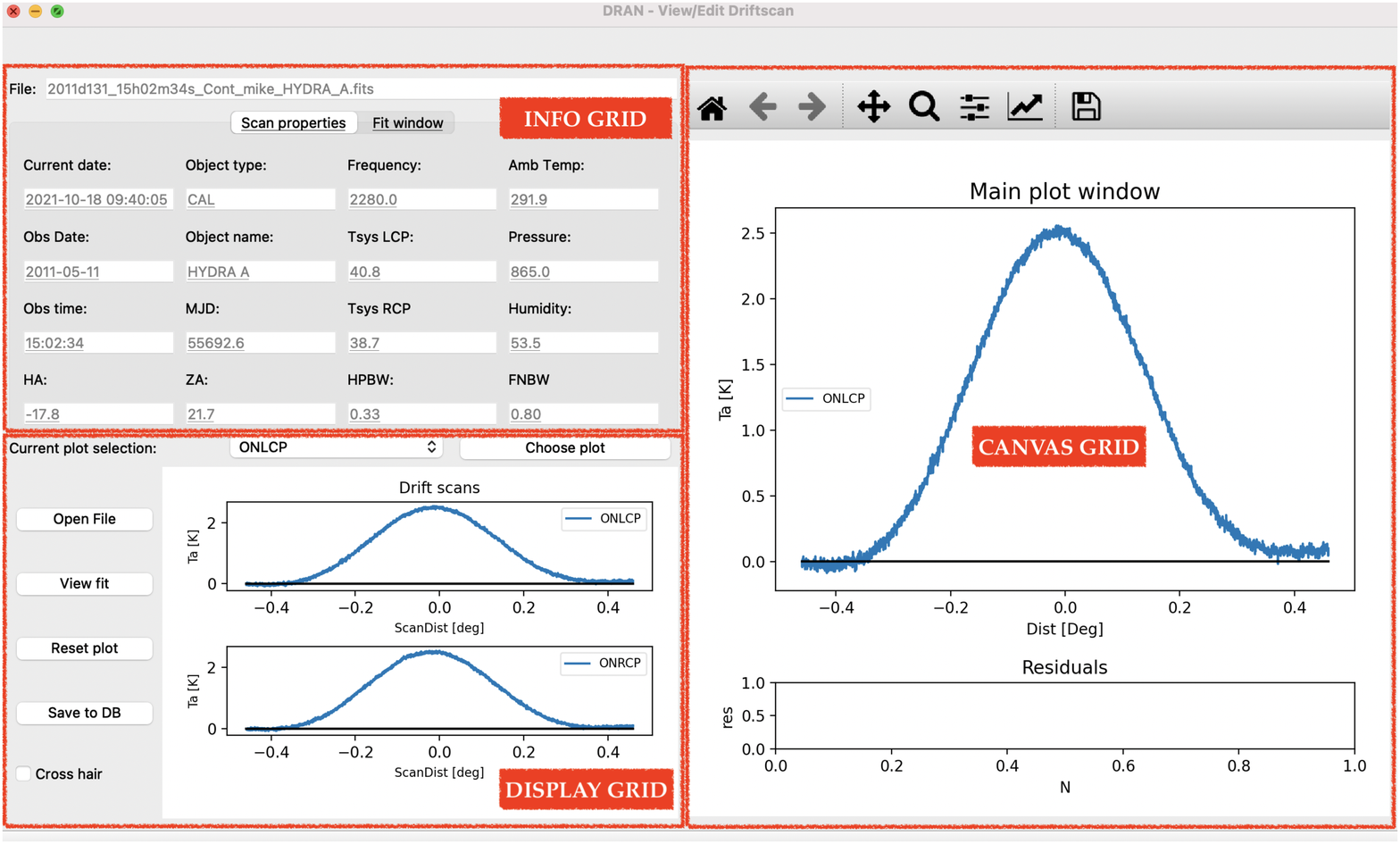}{ex_fig2}{Snapshot of the {\tt DRAN} graphical user interface.}

HartRAO observes several calibrators and target sources, so having an interactive interface is highly beneficial when one needs to run a quick analysis or simply make quick views of an observation before further analysis is carried out.  the {\tt DRAN} graphical user interface is very easy to use and provides a seamless platform on which to process HartRAO continuum drift-scan data.

\section{Future development}
Although {\tt DRAN} has had a significant improvement in the processing of HartRAO data, there is still a significant amount of work to be done. Future releases will include extra features of the program  like features to improve the manual operation of the program., adding better control on the analysis GUI and also improving the documentation HTML page to allow interactivity.

\bibliography{X3-023}


\end{document}